# Direct measurements of the magnetocaloric effect in ribbon samples of Heusler alloys Ni - Mn - M (M = In, Sn)

A. M. Aliev[a,*], A. B. Batdalov[a],
V. V. Koledov[b], V. G. Shavrov[b],
V. D. Buchelnikov[c],
J. García[d], V. M. Prida[d], B. Hernando[d]

[(a)] Amirkhanov Institute of Physycs of Daghestan Scientific Center, RAS, Makhachkala 367003, Russia

[(b)] Kotelnikov Institute of Radio Engineering and Electronics, RAS, Moscow 125009, Russia

[(c)] Chelyabinsk State University, 454001 Chelyabinsk, Russia

[(d)] Depto. de Física, Facultad de Ciencias, Universidad de Oviedo, Calvo Sotelo s/n, 33007 Oviedo, Spain

Direct measurements of the magnetocaloric effect in samples of rapidly quenched ribbons of $Mn_{50}Ni_{40}In_{10}$ and $Ni_{50}Mn_{37}Sn_{13}$ Heusler alloys, with potential applications in magnetic refrigeration technology, are carried out. The measurements were made by a precise method based on the measurement of the oscillation amplitude of the temperature in the sample while is subjected to a modulated magnetic field. In the studied compositions both direct and inverse magnetocaloric effects associated with magnetic (paramagnet - ferromagnet - antiferromagnet) and structural (austenite - martensite) phase transitions are found. Additional inverse magnetocaloric effects of small value are observed around the ferromagnetic transitions.

e-mail: lowtemp@mail.ru

In recent years the magnetic compounds with significant magnetocaloric effect (MCE) near room temperature are intensively studied. On the basis of these studies the possibility of ecological, economically viable solid-state refrigerators is proposed [1]. Recently, among the promising materials for magnetic refrigeration, a considerable interest is attracted to the Heusler alloys family Ni-Mn-X (X = Ga, Sn, Sb, In), in which a giant MCE is found [2, 3]. These alloys are characterized by the fact that near room temperature they exhibit a sequence of magnetic (paramagnetic - ferromagnetic - antiferromagnetic) and structural (austenite-martensite) phase transitions. In some cases, the magnetic and structural transitions merge into magnetostructural transitions. In the region of second order PM-FM phase transition, near $T_C$, the direct MCE, while at the martensitic transition of first order and near the metamagnetic transition of the FM-AFM, as well as merged metamagnetostructural transition, the inverse MCE is observed. The inverse MCE in bulk Heusler alloys samples is studied in [2-6].

In the most of MCE studies, indirect methods based on the measurement of isothermal magnetization curves are used. However, in many cases the use of indirect methods for obtaining data on magnetocaloric properties of the materials is insufficiently substantiated. This particularly concerns to the cases of first order magnetic phase transitions and magnetostructural transitions [7-9]. Therefore, it needs to develop direct methods for studying magnetocaloric properties, with high sensitivity, simplicity and adaptability.

In the present work, the MCE is studied in melt spun ribbons Ni-Mn-M (M = Sn, In) Heusler alloys by the direct technique, in modulating magnetic fields of small-amplitude, as it has been proposed in [10]. The use of the magnetocaloric materials in the form of thin films or ribbons in the manufacture of refrigerators can optimize the heat transfer between the working body and heat-exchange fluid, and thus improving the technical characteristics of refrigeration unit. However, it is difficult to investigate MCE in small samples by indirect methods, and the classical direct measurements are impossible to be conducted. The proposed method allows a high precision investigation of the magnetocaloric properties of samples with small size and mass performed in weak magnetic fields. The measurements were performed at frequencies ranging 0.3-0.5 Hz. The magnetic field during the experiment was always directed along the plane of the sample. The heat capacity was measured by an AC-calorimeter. Gd was used as the test sample for the calibration of the set-up.

With the help of the technique, we have investigated the MCE in $Ni_{50}Mn_{37}Sn_{13}$ and $Mn_{50}Ni_{40}In_{10}$ Heusler alloys. Samples with typical dimensions 1x3x0.015 $mm^3$ were cut from ribbons obtained by rapid quenching from the melt. Ribbons have a textured microcrystalline structure, with elongated column grains perpendicularly oriented to the ribbon plane [11, 12].

According to [12], the $Mn_{50}Ni_{40}In_{10}$ sample transforms at 311 K into the ferromagnetic state. The martensitic phase transition starting and finishing temperatures are $M_S$ = 213 K, $M_f$ = 173 K, respectively, whereas the corresponding ones to the austenite transition are $A_S$ = 222 K, and $A_f$ = 243 K, respectively. The $Ni_{50}Mn_{37}Sn_{13}$ sample transforms into the ferromagnetic state at $T_C$ = 313 K, and the martensitic transformation and austenite transition temperatures are $M_S$ = 218 K, $M_f$ = 207 K and $A_S$ = 224 K, $A_f$ = 232 K, respectively [11].

It is worthwhile mentioning that characteristic features of magnetostructural transitions in Heusler alloys quite clearly reveal themselves in the magnetization and susceptibility behavior under weak magnetic fields [11 - 13]. At the same time, the MCE study is usually carried out in strong magnetic fields, when the mentioned features are suppressed. Concerning with this, the MCE investigation in weak magnetic fields is of particular interest for studying the nature of phase transitions in Heusler alloys.

Fig. 1 shows the temperature dependence of the specific heat for $Ni_{50}Mn_{37}Sn_{13}$ in a zero magnetic field and in the field of 11 kOe. High-temperature specific heat anomaly displays a maximum at $T$ = 309.5 K, corresponding to the FM-PM phase transition, which takes place in the austenite phase. At temperatures of 217 K (cooling mode) and 233 K (heating mode) a second anomaly in the heat capacity corresponding to a first order metamagnetostructural phase transition is observed. The heat jump $\Delta C$ is different for both regimes and correlates with the magnetization behavior [11]. The presence of strongly pronounced hysteresis ($\Delta T$ = 16K) indicates the phase transition as a first order one.

Fig. 2 shows the temperature dependence of MCE in $Ni_{50}Mn_{37}Sn_{13}$ alloy measured with an amplitude of $\Delta H$ = 500 Oe for the modulated magnetic field. MCE shows a maximum at $T$ = 312.5 K, near $T_C$, while around the martensitic transition temperature – the inverse MCE exhibits two maxima: around $T$ = 224 K (cooling mode) and at $T$ = 236 K (heating mode). In addition, in the temperature range 284-302 K, below the ferromagnetic transition there is another peak of inverse MCE with a maximum at $T$ = 297 K. All these features indicate the existence of a rather complicated picture of the MCE. As it is shown in [12] the temperature dependence of the low-field magnetic susceptibility of melt spun ribbons has two peaks. Such magnetization behavior can be explained as follows. As is known, the low-field susceptibility is caused by process of domain walls displacements [14]. Susceptibility is directly proportional to the saturation magnetization and inversely proportional to the anisotropy of the magnetic substance [14, 15]. When approaching the Curie point the degree of reduction of the anisotropy intensity is higher than the corresponding to the saturation magnetization. In this regard, the susceptibility near Curie point shows a peak (Hopkinson effect). In the low-field range, magnetization displays a second peak that can be explained by the model proposed in [16] for Heusler alloys, where there is a possibility of the sign inversion of the exchange interaction. According to [16], in Heusler alloys, the phase transition from ferromagnetic phase to the antiferromagnetic (or paramagnetic) phase takes place accompanied by structural changes. In this case, a decrease in magnetization at the transition from cubic austenite phase to the martensitic phase is due to the simultaneous FM-AFM transition. It is also possible, that such a decrease in magnetization may be due to the fact that in Ni-Mn-X Heusler alloys the structural phase transition that is observed at low temperatures is accompanied by the origin of AFM ordering of manganese atoms in the ferromagnetic matrix (i.e. the mixed ferro-antiferromagnetic phase) in the martensitic phase [2].

Clearer picture of direct and inverse MCE is observed in $Mn_{50}Ni_{40}In_{10}$ (Fig. 3). Near $T_C$, direct MCE peak at $T$ = 314 K, and around the martensitic phase transition temperatures two inverse MCE peaks are observed. As can be seen, similarly to the case for $Ni_{50}Mn_{37}Sn_{13}$, when the temperature goes down, in the austenite phase near the PM-FM transition there is a direct MCE. At further cooling the first order structural austenite - martensite phase transition is observed with the second negative peak ($T_M$ = 201K), which is a consequence of the appearance of antiferromagnetic interactions in the martensite phase. In heating mode, the maximum of inverse MCE is shifted toward higher temperatures ($T_M$ = 237K) in accordance with the structural transition temperature shift. Inverse MCE is manifested at the structural phase transition austenite - martensite, due to the coexistence in the martensitic phase of FM and AFM interactions between manganese atoms [2, 15, 17]. The peak value of ΔT at the direct martensitic transformation temperature, $\Delta T_{cooling}$ = 0.00166 K (for magnetic field change $\Delta H$=370 Oe), is less than the value at the reverse transformation, $\Delta T_{heating}$=0.0042 K. This difference was to be expected because the magnetization change rate is larger upon the reverse martensitic transformation (insert in Fig. 3), though the magnetization change $\Delta M$ is larger upon the direct transformation [5].

Fig. 3 shows that near the magnetic phase transition in the austenite phase there is a peak of inverse MCE, despite the absence of any features on the low-field magnetization in $Mn_{50}Ni_{40}In_{10}$ ribbons at this temperature range [12]. The observed inverse MCE may be probably due to an additional phase shift, as a consequence of the presence in the austenitic phase, of both ferromagnetic and antiferromagnetic exchange, simultaneously. Such antiferromagnetic exchange is due to the additional number of Mn atoms in the crystal structure in comparison with the stoichiometric composition $Ni_2MnX$ [2, 17]. To determine the nature of inverse MCE above the Curie temperature of austenitic phase further careful measurements of magnetic properties in $Mn_{50}In_{40}In_{10}$ melt spun ribbons are required.

Simple linear extrapolation of our low-fields data yields 0.24 K and 0.41 K change in temperature ΔT with 1 T magnetic field variation for $Mn_{50}Ni_{40}In_{10}$ and $Ni_{50}Mn_{37}Sn_{13}$ respectively. More realistic value of the MCE can be extract from a field dependence of MCE (Fig. 4). The experimental data can be fitted by expression $\Delta T=aH^n$ [1], where $n$=0.88. From this follow ΔT=0.16 K at field change of 1 T. As we can see, the magnitude of MCE in the melt spun ribbons is small. Nevertheless, melt spun ribbons can be effectively employed in magnetic cooling devices through more efficient heat transfer. One can expect that in magnetic refrigerators the thermodynamic efficiency will be greatly enhanced by increasing the frequency of cycles. It is expected that films and melt-spun ribbons having faster heat transfer as compared with bulk samples, would be preferable. Thus, Heusler alloys in ribbon-shape samples, having the optimal combination of thermodynamical and magnetic characteristics, can compete with other magnetic coolant materials.

In conclusion, this paper reports on the application of a MCE measurement method using modulated magnetic field, for the direct MCE investigation in samples of melt-spun Heusler alloys ribbons, with small mass and thickness. The findings point out the complex nature of structural and magnetic phase transition in these alloys, which needs further investigation.

**Acknowledgement**

This work was supported by RFBR (09-08-96533, 09-08-01177, 10-02-92662), FANI (02.513.12.3097), Research Program of DPS RAS, MAT2009-13108-C02-01 and FC09-IB09-131.

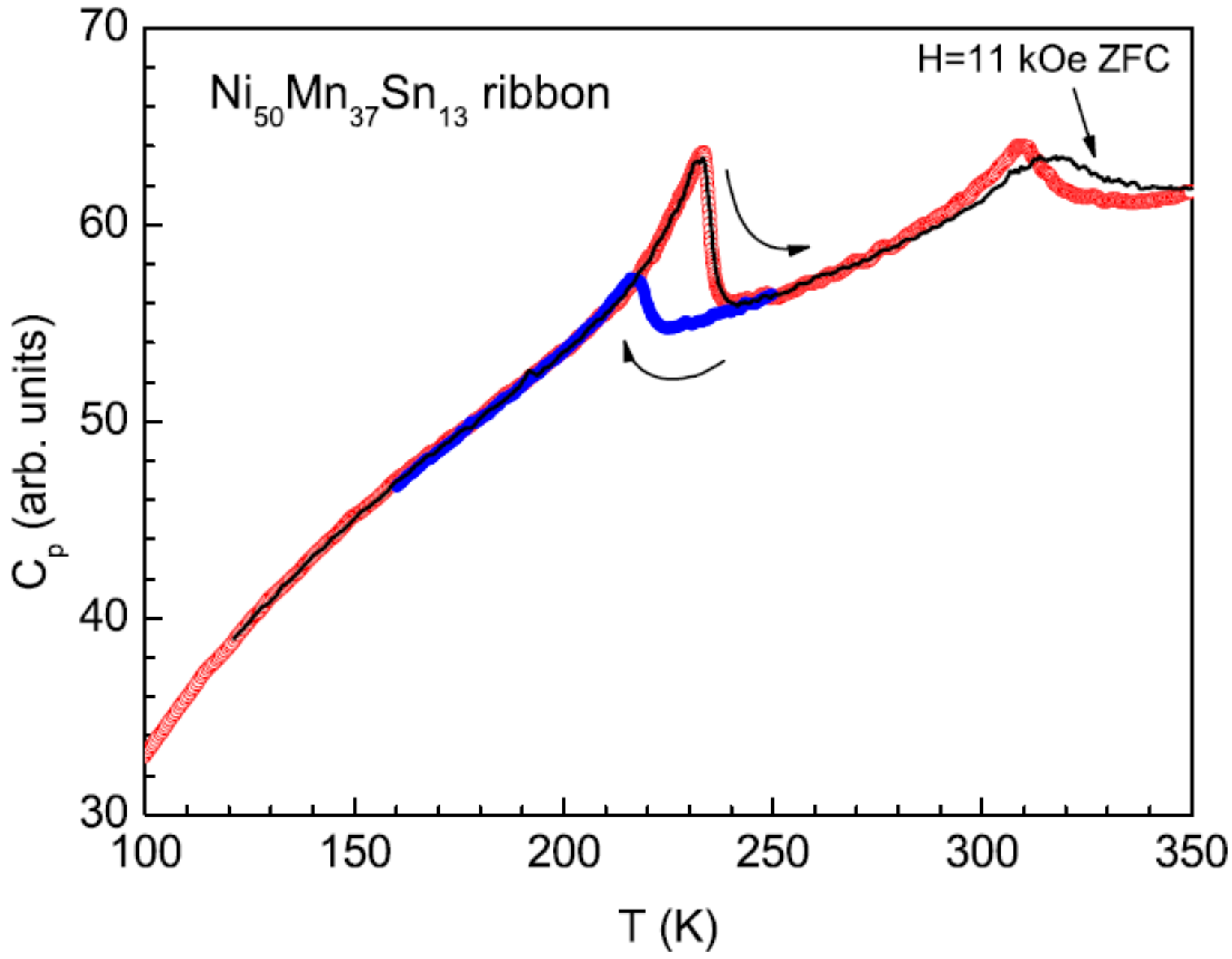

Figure 1. Specific heat of $Ni_{50}Mn_{37}Sn_{13}$ at heating and cooling and in magnetic field of 11 kOe.

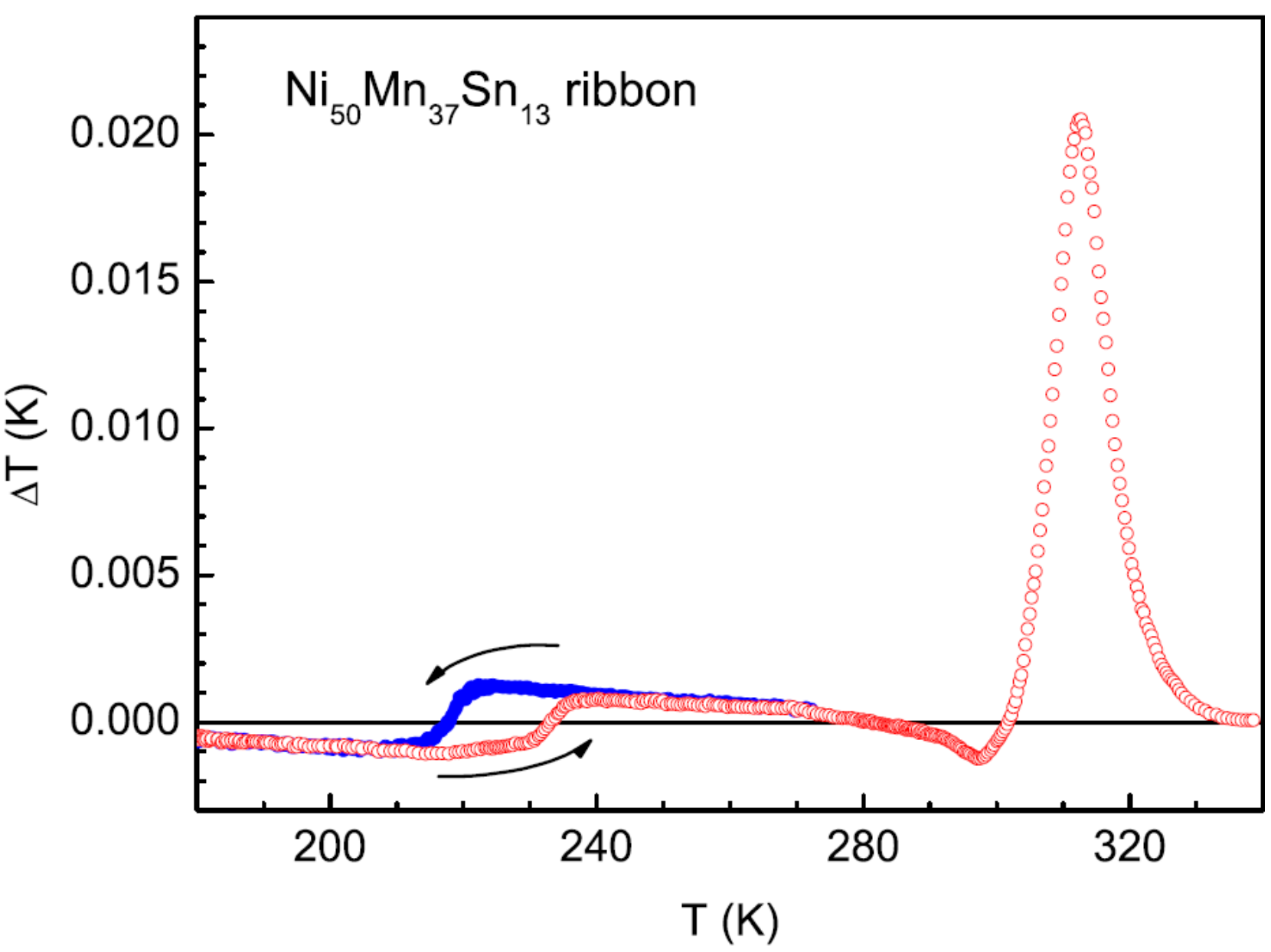

Figure 2. MCE in $Ni_{50}Mn_{37}Sn_{13}$ at heating and cooling at magnetic field change of 500 Oe.

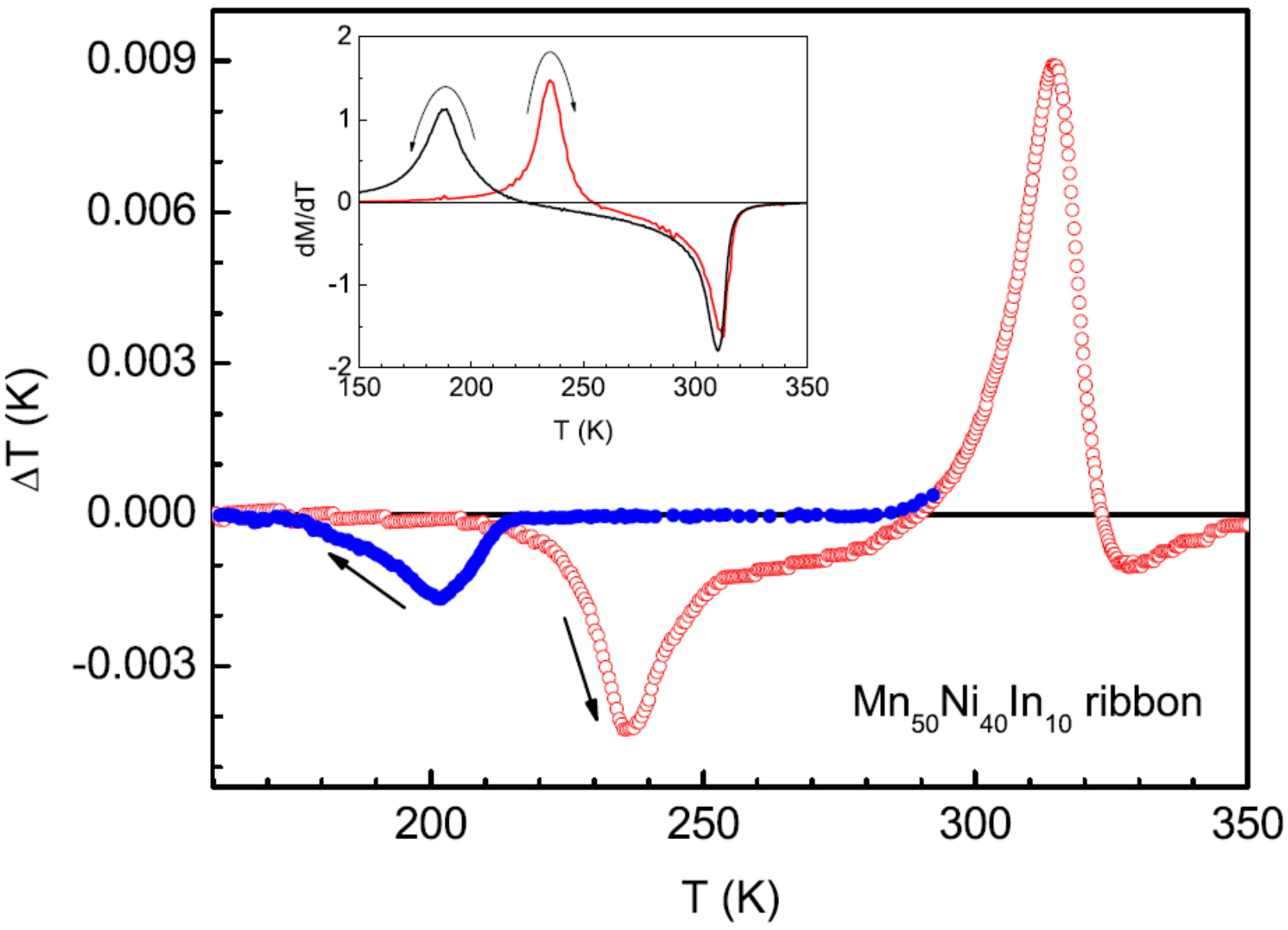

Figure 3. MCE in $Mn_{50}Ni_{40}In_{10}$ at heating and cooling at magnetic field change of 370 Oe. Inset – dM/dT at heating and cooling run.

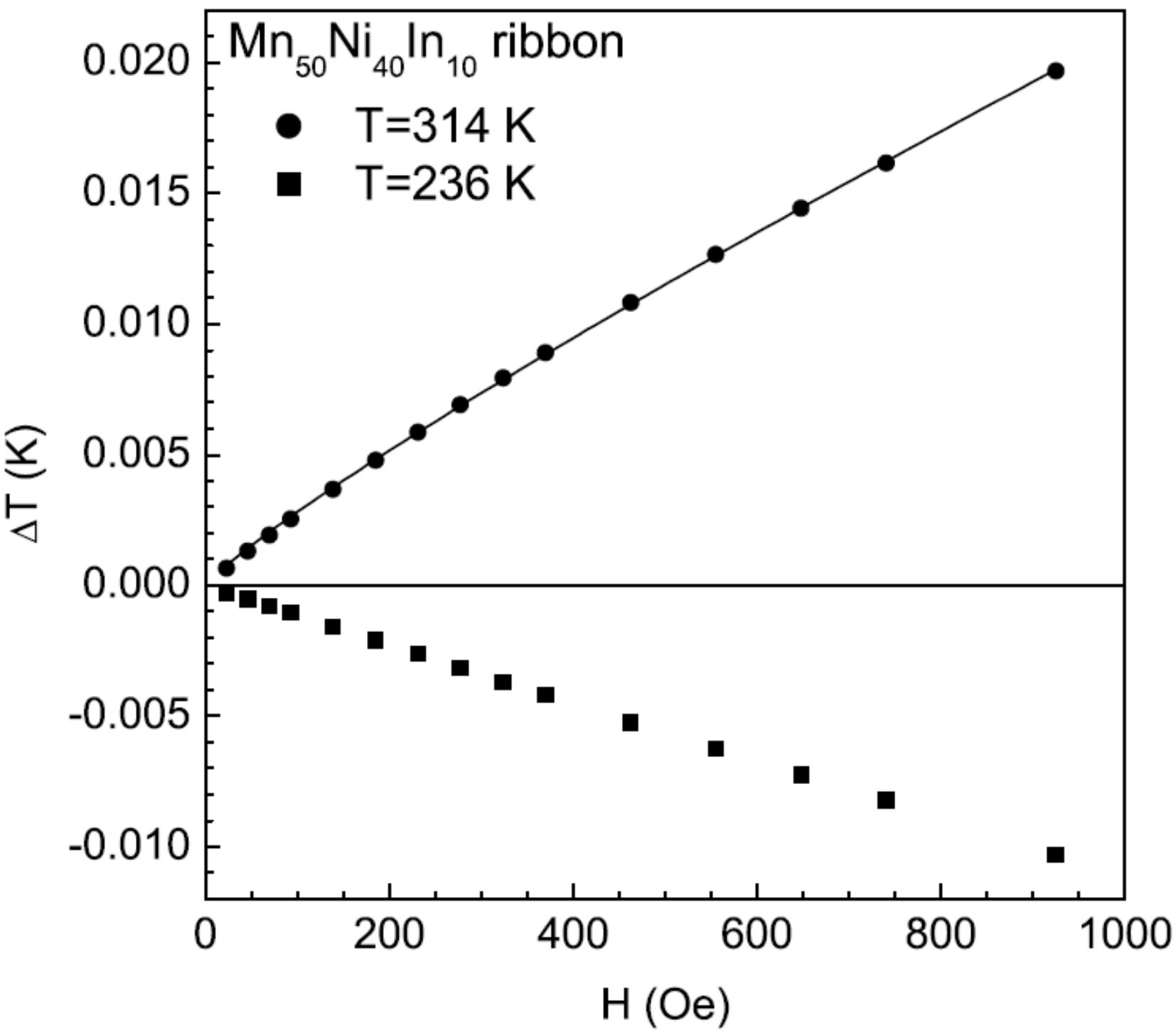

Figure 4. Field dependence of MCE in $Mn_{50}Ni_{40}In_{10}$.